\documentclass[journal]{IEEEtran}

\usepackage{amssymb}
\usepackage[cmex10]{amsmath}
\usepackage{stfloats}
\usepackage{graphicx}
\usepackage{subfigure}
\usepackage{tabularx}
\usepackage{epsfig,epsf,color,balance,cite}
\usepackage{verbatim}
\usepackage{url}
\usepackage{bm}
\usepackage{caption}
\usepackage{threeparttable}
\usepackage{diagbox}
\usepackage{multirow}

\usepackage{algorithm}
\usepackage{algorithmic}
\definecolor{myc1}{rgb}{0,0,1}

\begin{document}

\title{Channel Assignment in Uplink Wireless Communication using Machine Learning Approach}

\author{
\IEEEauthorblockN{
Guangyu Jia, Zhaohui Yang, Hak-Keung Lam, \IEEEmembership{Fellow, IEEE}, Jianfeng Shi, and Mohammad Shikh-Bahaei
}
\thanks{G. Jia, Z. Yang, H. Lam, and M. Shikh-Bahaei are with the Centre for Telecommunications Research, Department of Engineering, King's College London, WC2R 2LS, UK, Emails: guangyu.jia@kcl.ac.uk, yang.zhaohui@kcl.ac.uk, hak-keung.lam@kcl.ac.uk, m.sbahaei@kcl.ac.uk.}
\thanks{J. Shi is with the School of Electronic and Information Engineering, Nanjing University of Information Science and Technology, Nanjing 210096, China. Email: jianfengshi16@hotmail.com.}
}

\maketitle

\begin{abstract}
This letter investigates a channel assignment problem in uplink wireless communication systems. Our goal is to maximize the sum rate of all users subject to integer channel assignment constraints. A convex optimization based algorithm is provided to obtain the optimal channel assignment, where the closed-form solution is obtained in each step. Due to high computational complexity in the convex optimization based algorithm,  machine learning approaches are employed to obtain computational efficient solutions.  More specifically, the data are generated by using convex optimization based algorithm and the original problem is converted to a regression problem which is addressed by the integration of  convolutional neural networks (CNNs), feed-forward neural networks (FNNs), random forest and gated recurrent unit networks (GRUs). The results demonstrate that the machine learning method largely reduces the computation time with slightly compromising of prediction accuracy.
\end{abstract}

\begin{IEEEkeywords}
Resource allocation, convex optimization, machine learning, deep learning.
\end{IEEEkeywords}
\IEEEpeerreviewmaketitle
\vspace{-0.8em}
\section{Introduction}
Driven by the rapid development of advanced multimedia applications, next-generation wireless networks must support massive connectivity.
The connectivity always results in the integer optimization problem, which was solved by conventional  {algorithms mostly {operating} in an off-line manner with high computation complexity and {depending} largely on accurate channel state information \cite{di2016sub}.}
Machine learning tools  \cite{8755300,8752012,chen2019joint,yang2019energy,wang2019deep,shi2018adaptive}  can exploit big data
for wireless network state estimation and find the relationship between the {decision} variables and objective functions in an
online manner so as to reduce the computational complexity.

There are many research attentions of applying machine learning for solving integer optimization problems in wireless communication.
For integer user association problem, the authors in \cite{zappone2018user} investigated the use of deep learning
to perform user-cell association for sum-rate maximization in
massive multiple input multiple output (MIMO) networks.
For integer cache placement problem, the optimization of the caching
locations was {transformed to a grey-scale image-based problem and deep convolutional neural network (CNN) was adopted accordingly \cite{wang2019caching}.}
{Also, the linear sum assignment problems were solved through using the
	deep neural networks (DNNs) \cite{lee2018deep}. }
Moreover, in device-to-device (D2D) networks,
a novel graph embedding based method for link
scheduling {was proposed} \cite{lee2019graph}.
Further 
reinforcing the generalization ability in D2D networks,
a DNN structure was proposed in \cite{lee2019learning} with a novel loss function to achieve better
dynamic control over optimality and computational complexity.
However, the above contributions \cite{zappone2018user,wang2019caching,lee2018deep,lee2019graph,lee2019learning} all restricted that the number of users is assumed to be the same as wireless {resources}, which cannot meet the massive number of devices in future communication systems. \textcolor{black}{Besides, the machine learning approaches in  \cite{zappone2018user,wang2019caching,lee2018deep,lee2019graph,lee2019learning} are all neural networks and have the  limitations of prediction performance. To further investigate machine learning methods' capacity and potential in wireless communication domains, {this} letter employs  several machine learning algorithms as base learners and {integrates} them by {an} ensemble learning approach. By employing different machine learning methods, we aim to improve the diversity of base models and the final prediction performance as well as the model's generalization capability and robustness when exposed to contaminations. }

\textcolor{black}{
The contributions of this letter is summarised as follows:
\vspace{-0.1cm}
\begin{itemize}
	\item The channel assignment problem is formulated for the case that the number of users is larger than the number of subchannels. A convex optimization based algorithm is proposed to obtain the globally optimal channel assignment despite of the integer constraints.
	\item The convex optimization problem is converted to a regression problem and solved by machine learning frameworks, which yield rigorously optimal and computationally efficient solutions.
	\item Ensemble learning is utilized to combine different machine learning models and improve the {prediction} performance. Also, different combinations of optimization algorithms are adopted and compared.
	\item The computation time is largely reduced by the proposed machine learning frameworks without much compromising of prediction accuracy.
\end{itemize}}


\vspace{-0.75em}
\section{System Model and Problem Formulation}
Consider {an} uplink single cell network with $M$ users and $N$ subchannels.
Denote $\mathcal M=\{1, 2, \cdots, M\}$ as the set of all users  and $\mathcal N=\{1, 2, \cdots, N\}$ as the set of all subchannels.
Let $p_i$ denote the uplink transmission power for user $i$.
Binary variable $x_{ij}$ reflects the association relationship between user $i$ and subchannel $j$, i.e., $x_{ij}=1$ means that user $i$ occupies sbuchannel $j$; otherwise $x_{ij}=0$.

Due to massive number of users and limited number of channels, i.e., $M\geq N$, non-orthogonal multiple access (NOMA) scheme is utilized by serving multiple users in each subchannel \cite{7390209}.
With successful interference cancellation in NOMA\footnote{\textcolor{black}{The transmission power of each user is assumed to be fixed, thus the power constraints are not involved in the optimization problem.}},  the sum rate of all users occupying subchannel $j$ is given by \cite{7390209,8125101}:
\vspace{-0.5em}
\begin{equation}
\vspace{-0.5em}
r_{j}=B\log_2\left(1+\frac{\sum_{i=1}^M x_{ij}p_ih_{ij}}{\sigma^2B}
\right),
\end{equation}
where $B$ is the bandwidth of each subchannel,
$h_{ij}$ is the channel gain between user $i$ and the {base station (BS)} on subchannel $j$, and $\sigma^2$ is the power density of Gaussian noise.

The sum rate maximization problem for assigning users with subchannels can be formulated as:
 \begin{subequations}\label{sys1min1}\vspace{-0.5em}
\begin{align}
\mathop{\max}_{\boldsymbol x } \quad & \sum_{j=1}^N B\log_2\left(1+\frac{\sum_{i=1}^Mx_{ij}p_ih_{ij}}{\sigma^2B}
\right)\\
\textrm{s.t.}\quad
& \sum_{j=1}^N x_{ij}=1, \quad \forall i \in \mathcal M,\\
&  \sum_{i=1}^M x_{ij} = A, \quad \forall j \in \mathcal N,\\
&    x_{ij}\in\{0,1\}, \quad \forall i \in\mathcal M,j \in \mathcal N,
\end{align}
\end{subequations}
{where $\boldsymbol x=[x_{11},x_{12},\cdots,x_{MN}]^T$} and $A$ is the allowed number of associated users in one subchannel.
Constraints (\ref{sys1min1}b) ensue that each user only {occupies} one subchannel.
\vspace{-0.5em}
\section{Convex Optimization Based Algorithm}
\vspace{-0.5em}
Due to integer constraints (\ref{sys1min1}d), it is hard to solve linear integer problem.
By temporarily relaxing the integer constraints (\ref{sys1min1}d) with $x_{ij}\in[0,1]$, problem \eqref{sys1min1} is a linear problem, also convex.
For convex problem \eqref{sys1min1} with relaxed constraints, the optimal solution can be effectively obtained by using the dual method \cite{boyd2004convex}.

Let $\lambda_i$
denote the dual variable associated
with constraint $i$ in (\ref{sys1min1}b),
we obtain the dual problem of the problem in (\ref{sys1min1}), which is given by:
\vspace{-0.5em}
\begin{equation}\label{co2eq1}\vspace{-0.5em}
\mathop{\min}_{\boldsymbol{\lambda}}\quad D(\boldsymbol{\lambda}),
\end{equation}
where
\vspace{-0.5em}
\begin{equation}\label{co2eq1_2}\vspace{-0.5em}
D(\boldsymbol{\lambda})=\left\{ \begin{array}{ll}
\!\!\!\mathop{\max}\limits_{\boldsymbol x}\;\;
&  \mathcal L_1 (\boldsymbol x, \boldsymbol \lambda)
\\
\!\!\rm{s.\;t.}&  \sum_{i=1}^M x_{ij}=A, \quad \forall j \in \mathcal N,\\
&    x_{ij}\in[0,1], \quad \forall i\in\mathcal M,j \in \mathcal N,
\end{array} \right.
\end{equation}
\vspace{-0.5em}
\begin{align}\label{co2eq2}\vspace{-0.5em}
&\!\!\! \mathcal L_1(\boldsymbol x, \boldsymbol \lambda)\!=\!
\sum_{j=1}^N\! B\log_2\!\left(\!1\!+\!\frac{\sum_{i=1}^M\!x_{ij}p_ih_{ij}}{\sigma^2B}
\!\right)\!
\!+\!\sum_{i=1}^M \!\lambda_i\!\left(\!
\sum_{j=1}^N \! x_{ij} \!-\!1\!\right)\!
\nonumber\\=\!&
\sum_{j=1}^N \!B\log_2\left(\!1+\!\frac{\sum_{i=1}^Mx_{ij}p_ih_{ij}}{\sigma^2B}
\!\right)\!
\!+\!\sum_{j=1}^N \!\left(\!
\sum_{i=1}^M \lambda_i x_{ij} \!-\!\lambda_i\! \right)\!,\!\!\!\!\!
\end{align}
and $\boldsymbol \lambda=[\lambda_1,\cdots, \lambda_N]^T$.

To minimize the objective function in (\ref{co2eq1_2}), the optimal $x_{ij}$ can be calculated via the Karush-Kuhn-Tucker (KKT) conditions.
Since both the objective function and constraints can be decoupled into $N$ subchannels, the subproblem of (\ref{co2eq1_2}) in subchannel $j$  can be given by:
\vspace{-0.5em}
 \begin{subequations}\label{sys1min1_2}\vspace{-0.3em}
\begin{align}
\mathop{\max}_{\boldsymbol x_j } \quad & B\log_2\left(1+\frac{\sum_{i=1}^Mx_{ij}p_ih_{ij}}{\sigma^2B}
\right) + \sum_{i=1}^M \lambda_i x_{ij}
 \\
\textrm{s.t.}\quad
&  \sum_{i=1}^M x_{ij}=A, \\
&    x_{ij}\in[0,1], \quad \forall i \in \mathcal M,
\end{align}
\end{subequations}
where $\boldsymbol x_j=[x_{1j},x_{2j},\cdots,x_{Mj}]^T$.

It can be proved that (\ref{sys1min1_2}) is a convex problem,
and the Lagrangian function of problem (\ref{sys1min1_2}) is
\vspace{-0.5em}
\begin{align}\vspace{-0.5em}
\mathcal L_2&(\boldsymbol{x_j}, \alpha,\boldsymbol \beta, \boldsymbol \gamma)=
 B\log_2\left(1+\frac{\sum_{i=1}^Mx_{ij}p_ih_{ij}}{\sigma^2B}
\right) + \sum_{i=1}^M \lambda_i x_{ij}
\nonumber\\&
+\alpha\left(\sum_{i=1}^M x_{ij}-A\right)+\sum_{i=1}^M\beta_i x_{ij}
+\sum_{i=1}^M\gamma_i (1-x_{ij}).
\end{align}

The KKT conditions of problem (\ref{sys1min1_2}) are \cite{boyd2004convex}
\begin{subequations}\label{KKT1}
\begin{align}
\frac{p_ih_{ij}}
{(\ln2)\left(\sum_{i=1}^Mx_{ij}p_ih_{ij}+\sigma^2B\right)}\!+\! \lambda_i\!+\!\alpha\!+\!\beta_i\!-\!\gamma_i
&=0,  \\
\sum_{i=1}^M x_{ij}&=A,\\
\beta_i x_{ij}&=0,\\
\gamma_i (1-x_{ij})&=0.
\end{align}
\end{subequations}
\textcolor{black}{
	In the left hand side of  (\ref{KKT1}a), it is founded that the denominator of the first term is the same for every user $i$.
	With this finding, we introduce a new variable $\kappa$, which is defined as:
\vspace{-0.5em}
	\begin{equation}\label{co2eq3_1}\vspace{-0.5em}
	\kappa={(\ln2)\left(\sum_{i=1}^Mx_{ij}p_ih_{ij}+\sigma^2B\right)}.
	\end{equation}
	With the help of $\kappa$, we can directly obtain the optimal solution of $x_{ij}^*$ according to the following analysis.
	Substituting \eqref{co2eq3_1} to (\ref{KKT1}a) yields:
\vspace{-0.5em}
	\begin{equation}\label{co2eq3_0}\vspace{-0.5em}
	\frac{p_ih_{ij}   +\kappa\lambda_i}{\kappa}=\gamma_i-\beta_i-\alpha.
	\end{equation}
	Based on (\ref{KKT1}c)-(\ref{KKT1}d), we can obtain that (i) $x_{ij}=1$ if and only if $\gamma_i>0$ and $\beta_i=0$ and (ii) $x_{ij}=1$ if and only if $\gamma_i=0$ and $\beta_i\geq0$, which indicates that the right hand side of \eqref{co2eq3_0} should be as large as possible for the case $x_{ij}=1$.
	Besides, there are only $A$ users such that $x_{ij}=1$ according to (\ref{KKT1}b).
	Thus, the user with $x_{ij}=1$ should has the $A$ highest values of $\frac{p_ih_{ij}   +\kappa\lambda_i}{\kappa}$, i.e.,
\vspace{-0.5em}
	\begin{equation}\label{co2eq3}\vspace{-0.5em}
	x_{ij}^*(\kappa)=\left\{ \begin{array}{ll}
	\!\!1, &\text{if}\; i \in \mathcal M_{A}\\
	\!\!0, &\text{otherwise},
	\end{array} \right.
	\end{equation}
	where set $\mathcal M_A$ means the index of $A$ users with the top $A$ highest values in $p_ih_{ij}   +\kappa\lambda_i$.
	Equation \eqref{co2eq3} indicates that the optimal $x_{ij}^*$ is only a function of $\kappa$.
	To calculate the value of $\kappa$, we combine \eqref{co2eq3_1} and \eqref{co2eq3}, which yields:
\vspace{-0.5em}
	\begin{equation}\label{co2eq3_2}\vspace{-0.5em}
	\kappa={(\ln2)\left(\sum_{i=1}^N x_{ij}^*(\kappa)  p_ih_{ij}+\sigma^2B\right)}.
	\end{equation}
	With the increase of $\kappa$, the user with lower value in $p_ih_{ij}$ can have higher value in $p_ih_{ij}   +\kappa\lambda_i$, i.e., the value of $\sum_{i=1}^N x_{ij}^*(\kappa)  p_ih_{ij}$ decreases, which indicates that the right hand side of \eqref{co2eq3_2} decreases with $\kappa$.
	Since the left hand side of \eqref{co2eq3_2} increases with $\kappa$ and the right hand side of \eqref{co2eq3_2} decreases with $\kappa$,
	the unique value of $\kappa$ can be obtained by the bisection method.}

\begin{figure*}
	\centering
	\vspace{-2em}
	\includegraphics[width=7in]{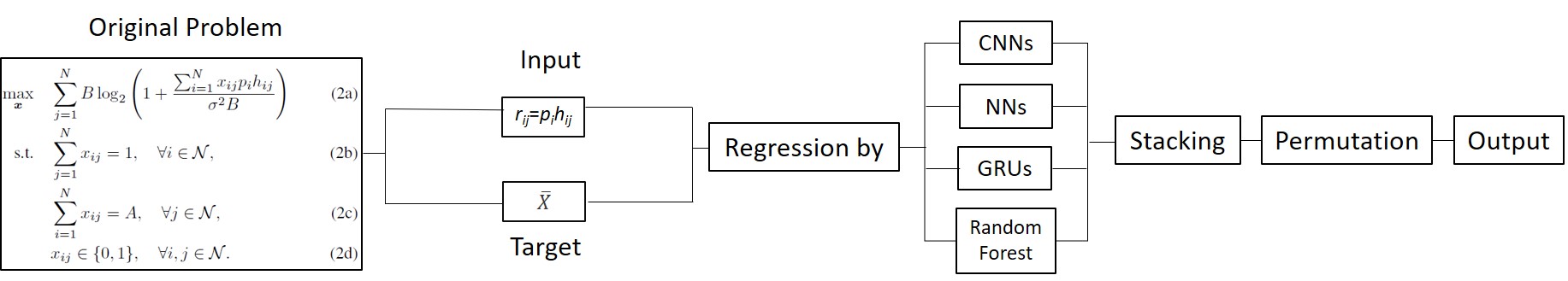}
\vspace{-0.5em}
	\caption{Block diagram of the implementation procedure in this letter.}
\vspace{-1.5em}
	\label{fig:overall}
\end{figure*}

Given $\boldsymbol x$, the value of $\boldsymbol \lambda$ can be determined by the gradient method \cite{bertsekas2009convex}.
By iteratively optimizing primal variable $\boldsymbol x$ and  dual variable $\boldsymbol \lambda$, the optimal subcahnnel allocation is obtained.
The dual method used to obtain the optimal subcahnnel allocation is given in Algorithm 1.
Notice that the optimal $x_{ij}$ is either 0 or 1 according to \eqref{co2eq3}, even though
we {relax $x_{ij}$} as continuous constraint in (\ref{sys1min1}). Therefore, we
can obtain optimal solution to problem (\ref{sys1min1}).

 \begin{algorithm}[h]
\caption{  Dual Method for Subchannel Association }
\begin{algorithmic}[1]
\STATE Initialize dual variable $\boldsymbol \lambda$.
\REPEAT
\STATE Update subchannel allocation $\boldsymbol x$  according to (\ref{co2eq3}).
\STATE Update dual variable  $\boldsymbol\lambda$ by the gradient method.
\UNTIL the objective value (\ref{sys1min1}a) converges
\end{algorithmic}
\end{algorithm}

{The major complexity in Algorithm 1} lies in updating subchannel allocation $\boldsymbol x$.
Based on \eqref{co2eq3}, the complexity of calculating $\boldsymbol x$ is
$\mathcal O\left(NM\log_2(1/\epsilon)\right)$, where $\mathcal O\left(\log_2(1/\epsilon)\right)$ is the complexity  of solving \eqref{co2eq3_2} by using the bisection method with accuracy $\epsilon$.
As a result, the total complexity of Algorithm 1 is $\mathcal O\left(LNM\log_2(1/\epsilon) \right)$, where $L$ is the number of iterations in Algorithm 1.

\vspace{-.5em}
\section{Deep Learning Based Approaches}\vspace{-.5em}
\textcolor{black}{In this letter, the machine learning approaches {are also involved to} obtain computational efficient solutions for the optimization problem.   The overall implementation procedures are shown in Fig. \ref{fig:overall}. The key of machine learning methods is to learn the relationship between {the constant input parameters $R = \{r_{ij} = p_i h_{ij}\}, i\in\mathcal M,j \in \mathcal N$} and the subchannel arrangements $X = \{x_{ij}\}, i\in\mathcal M,j \in \mathcal N$. To make the problem solvable, we convert the target one-hot matrix $X = \{x_{ij}\}, i\in\mathcal M,j \in \mathcal N$ to vector form $\bar{X} = \{\bar{x}_{i}\in \mathcal N\}, i \in \mathcal M$ so that the problem can be considered as a regression problem. Each column in $X$, i.e.,  $\{x_{i1},x_{i2},...,x_{iN}\}^{T}, i \in \mathcal M$ is {transformed} to an integer $\bar{x}_{i} \in \mathcal N, i \in \mathcal M$ which denotes the index of non-zero value (i.e., 1) in the each column, meaning the index of user in the $j$-th subchannel, i.e., $x_{i\bar{x}_{i}}=1$.}

After obtaining the input and target, the optimization problem is converted to a regression problem which can be modelled by machine learning frameworks. The adopted methods for regression consist of {CNNs, feed-forward neural networks (FNNs), random forest and gated recurrent unit networks (GRUs)}. In addition, ensemble learning method is utilized to combine different models and further improve the regression performance. After regression, the output of machine learning models are permutated according to the output values, thus, the integer outputs which represent the indices of users are obtained in the end.
\vspace{-.5em}
\subsection{Data Information}\vspace{-.5em}

\subsubsection{Data generation}
We deploy $M$ users uniformly in a square area of size $500$ m $\times$ $500$~m with the BS located at its center.
In each independent run, the channel gains of users on the subcahnnels are determined, i.e., $R$ is obtained. The output $X$ is accordingly calculated by using Algorithm 1. {The total number of generated samples is 1 million.}

\subsubsection{Data splitting}
The generated data are splitted to training, validation and test datasets. 70\% of the data are used for training while 10\% for test. 20\% of data are held out for validation which plays an important role in model ensemble stage, i.e., the stacking procedure.
\vspace{-.5em}
\subsection{Machine Learning Methods}\vspace{-.5em}
\textcolor{black}{Different machine learning methods are employed to solve the regression of the input $R$ and the target $\bar{X}$.} Ensemble learning is then utilized to make different models complement each other and improve the regression performance. \textcolor{black}{Different base learners are used as they can render diversity of the whole structure, which is beneficial for models' prediction accuracy, robustness and generalization capability when exposed to contaminations. The models are selected mainly through referencing published literatures of the successful models on similar datasets and trial and error of existed models. After comparing the performance of existed machine learning models, FNNs, CNNs, random forest and GRUs are employed as the base learners. The configurations of adopted models are shown in Table \ref{tabel:specification}, which are determined mainly through trial and error and taking into consideration of the input size. }

\begin{table*}
	\vspace {-1cm}
	\textcolor{black}{
	\footnotesize
	\renewcommand \arraystretch{1.5} 
	\renewcommand\tabcolsep{2pt}	
	\caption{\textcolor{black}{{Machine Learning Models' Specifications.} }}
	\begin{center}
		\begin{threeparttable}\scriptsize
			\begin{tabular}{p{0.4\columnwidth}|p{0.4\columnwidth}|p{0.4\columnwidth}|p{0.4\columnwidth}}\hline
				 \centering{FNNs} & \centering{CNNs} & \centering{GRU} &\quad\quad Random Forest \\ \hline
				 {Structure: basic FNN structure is shown in Fig. \ref{fig:nn}. Three hidden layers used in this letter are of the size  ([10, 10, 10], [10, 10, 10, 10], [20, 20, 20]); Optimizer: LM$^*$; Epoch: 300; Loss function: MSE$^{**}$}.  & Structure: basic CNN structure is shown in Fig. \ref{fig:cnn}. Layers' depth and filter size vary according to the input size; Optimizer: Adam + SGD with momentum; Batchsize: 500; Epoch: 40--60; Loss function: MSE. & Structure: Two GRU layers with 50 units each, followed by dropout layer and 4 Dense layers; Optimizer: Adam; Batchsize: 200; Epoch: 40; Loss function: MSE. & The number of trees in forest: 500; Criterion: Gini Impurity; The minimum number of samples required to split an internal node and to be at a leaf node: 2, 1, respectively. \\  \hline
			\end{tabular}
			\begin{tablenotes}
				\footnotesize
				\item[*]LM: Levenberg-Marquardt method.
				\item[**]MSE: mean squared error.
			\end{tablenotes}
		\end{threeparttable}
	\end{center}
	\label{tabel:specification}}
\vspace {-0.6cm}
\end{table*}	

\subsubsection{Deep Learning Models}
\begin{figure}
	\centering
	\includegraphics[width=8cm]{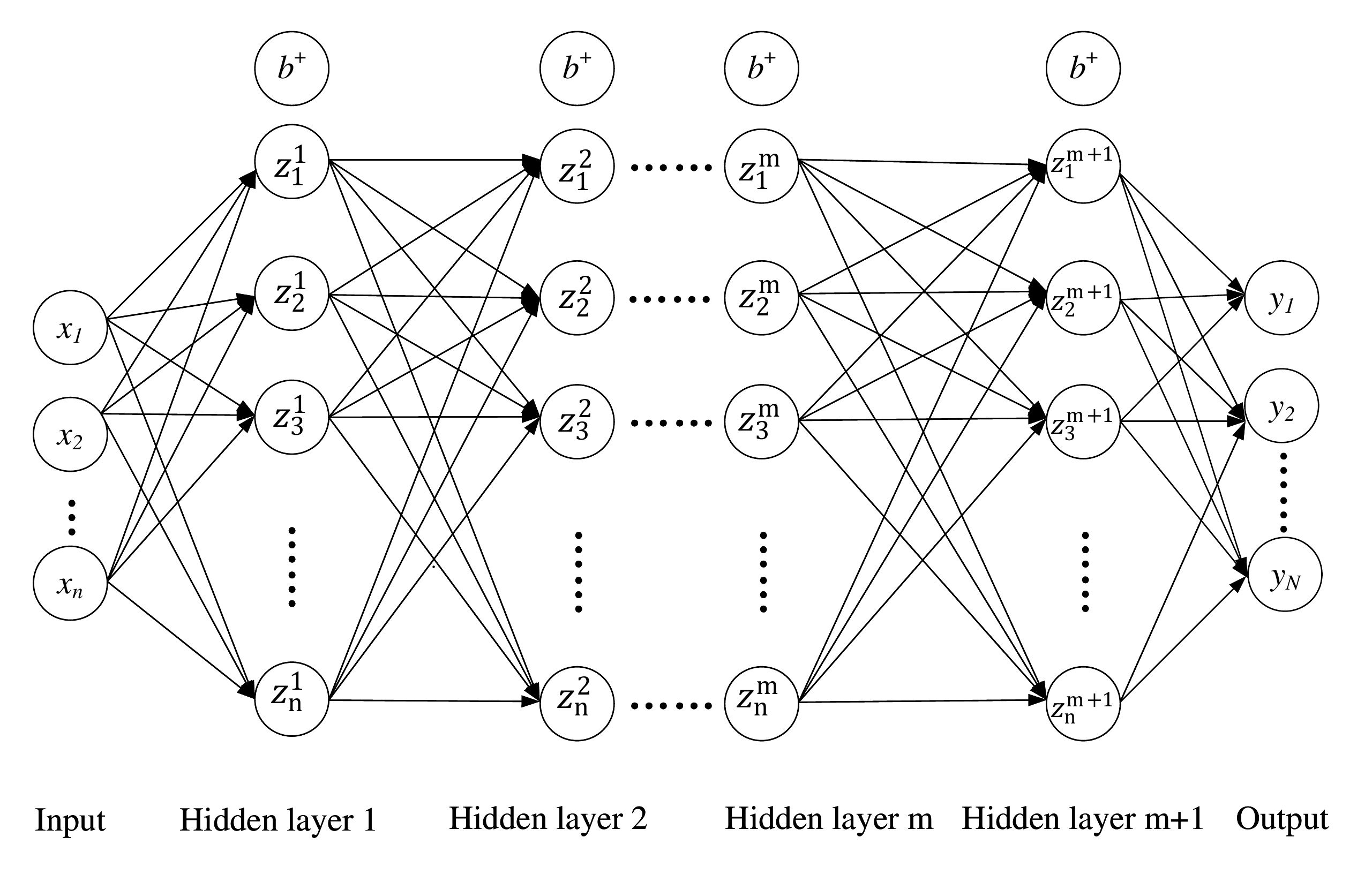}
\vspace{-0.5em}
	\caption{The structure of FNNs.}
\vspace{-0.5em}
	\label{fig:nn}
\vspace{-0.5em}
\end{figure}

FNNs are widely used in function approximation, pattern recognition and data classification. Due to its superiority in data fitting problems, we adopted FNNs as one of models to solve the regression problem. The general structure of FNNs used in this letter is shown in Fig.  \ref{fig:nn}.


\begin{figure*}
	\centering
	\hspace{-0.5cm}
	\includegraphics[width=15cm]{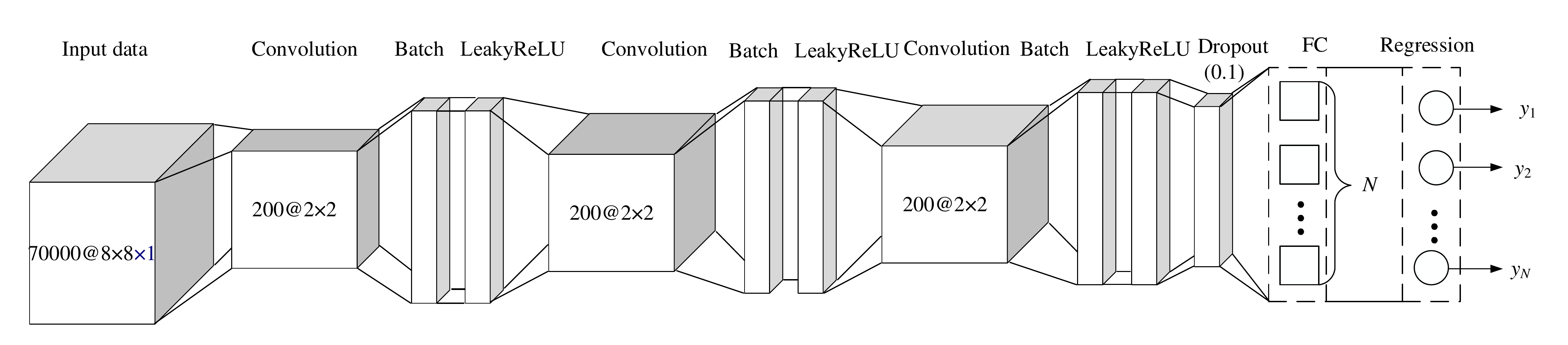}
	\caption{One CNN structure used in this letter.}
	\label{fig:cnn}
	\vspace{-1.5em}
\end{figure*}

CNNs have advantages in processing the high-dimensional data, such as images and time-series data due to (a) its convolutional setting in which the hidden units are not fully connected to the input but instead divided into locally connected segments; (b) pooling methods which reduce the dimensionality of the feature space and achieve invariance to small local distortions \cite{langkvist2014review}. One CNN architecture used in this letter is shown in Fig. \ref{fig:cnn}.

GRU architectures have superiority in capturing dependences of different time scales.
GRUs are similar to {Long Short-Term Memories (LSTMs)} which have gating units to adjust the information flow inside the unit but do not have a seperate memory cells \cite{chung2014empirical} and thus are faster to {be trained} than LSTMs but on a par with LSTMs' performance. When applying GRUs, the input data $R = \{r_{ij}\}, i,j \in \mathcal N$ are reshaped to vectors so as to form {the sequence inputs} for GRU networks. \textcolor{black}{As the correlation of channel model is based on the fact that the channel gain of each user is correlated in time domain \cite{1204012},  the GRU model is selected accordingly as one of the base learners.  }


\begin{figure}
	\centering
	\includegraphics[width=8.5cm]{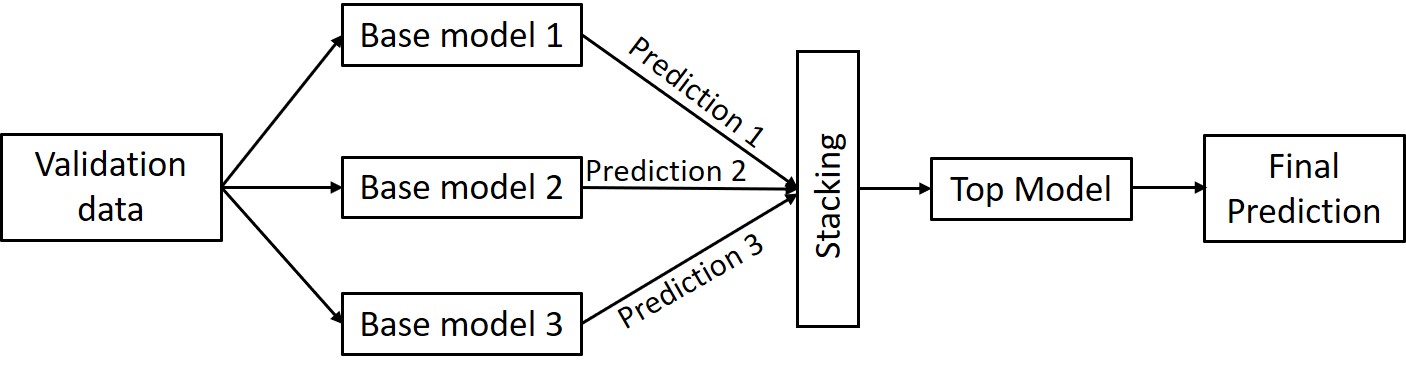}
	\caption{Stacking process.}
	\label{fig:meta}
\end{figure}
Stacking is a widely used ensemble method to boost the prediction performance \cite{wolpert1992stacked} of base learners. In this letter, we applied stacking method to a holdout dataset, i.e., the validation dataset, the predictions of base learners fed by validation data are then stacked to form a new dataset for training the top model. The stacking structure used in this letter is shown in Fig. \ref{fig:meta}. \textcolor{black}{The complexity of base learners and top models is no more than $\mathcal{O}(N^2)$ \cite{lee2018deep}.} The details of base models and top models in each case will be illustrated in the following section.

\subsubsection{Training method}

\textcolor{black}{The optimization {algorithm} utilized for CNNs' training in this letter is the combination of Adam and stochastic gradient descent (SGD) with momentum. In implementation, Adam is used first to realize fast gradient decent, then SGD is used for further optimizing the parameters.}

\begin{figure}
	\centering
	\includegraphics[width=8.5cm]{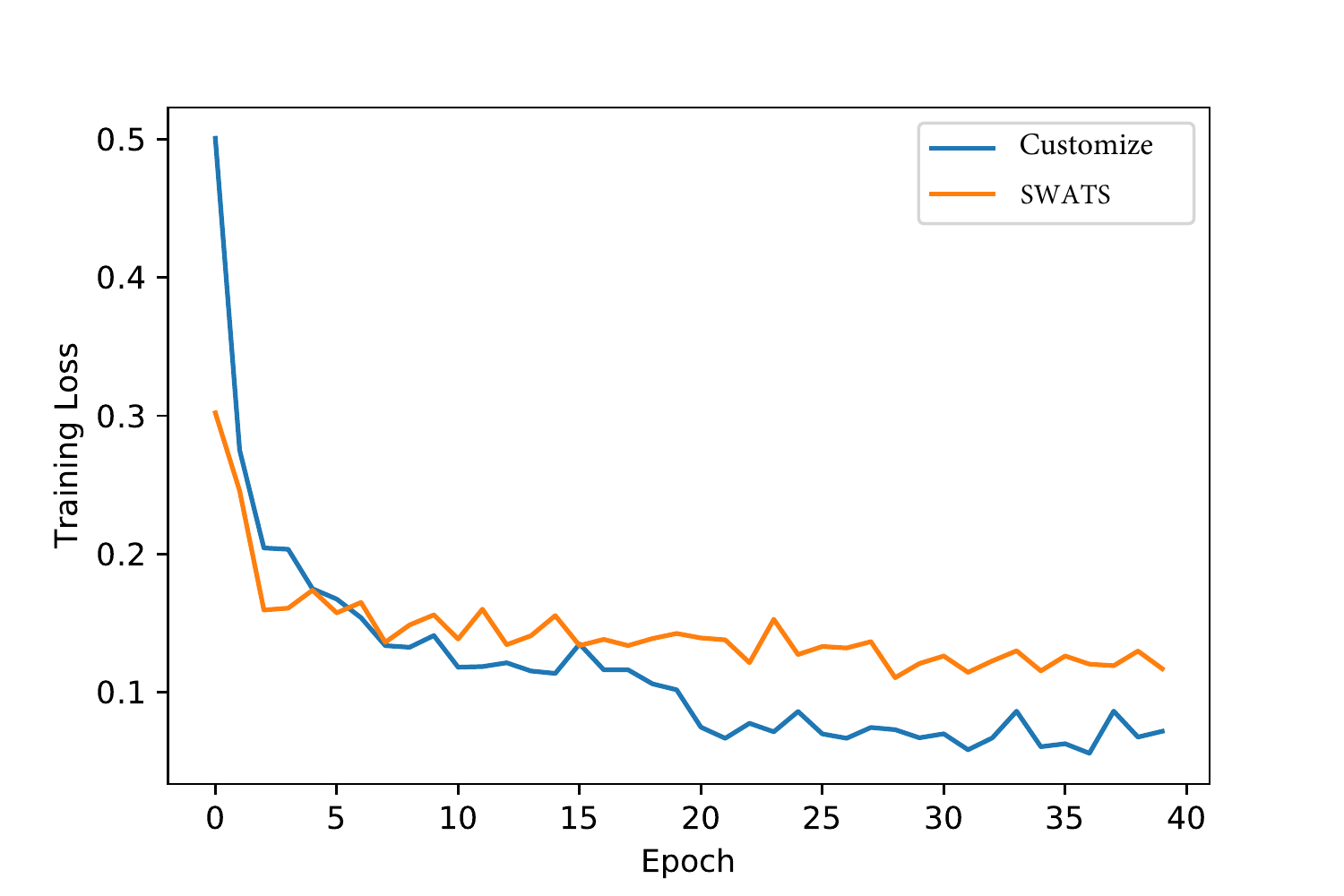}
	\caption{\textcolor{black}{Training loss of two training methods. SWATS is the method proposed in \cite{keskar2017improving}, Customize refers to the combination of Adam and SGD that the first half of training adopts Adam while the second half uses SGD algorithm.}}
	\label{fig:loss}
\end{figure}

\textcolor{black}{As shown in Fig. \ref{fig:loss}, the conversion from Adam to SGD is implemented in two ways: i) SWATS: adopting the SWATS strategy proposed in \cite{keskar2017improving} which automates the process of switching over by determining both the switchover point and the learning rate of SGD; ii) Customize: using Adam in the first half of training, i.e., the first 20 epochs of the example shown in Fig. \ref{fig:loss}; then use SGD in the second half of training, i.e., the last 20 epochs. From Fig. \ref{fig:loss}, it can be seen that at epoch 20 (the switchover point of the Customize method), there occurs obvious decrease of training loss from Adam to SGD. These two methods are compared in each case and the one having the better performance is employed in this letter.}
\vspace{-0.6em}
\section{Experiment Configuration}\vspace{-.3em}
This section will introduce the experiment procedure, platforms and the methodology for calculating the final accuracy.
The experiments were implemented in Matlab R2019a, Tensorflow 1.13.1 and Nvidia
GPUs. Matlab is for training FNNs with {Levenberg-Marquardt (LM)} algorithm, other classifiers were trained
in Tensorflow and sped up by Nvidia P100/GTX 1050 GPUs.
\vspace{-1em}
\subsection{Experiment Procedures}\vspace{-.3em}
After transforming the optimization problem to a regression problem, the machine learning methods including FNNs, CNNs, {GRUs}, \textcolor{black}{random forest} as well as the stacking method are employed to learn the pattern of channel assignment.  \textcolor{black}{Two type of cases were considered in this letter: 1) $M = N,$ meaning the same number of users and subchannels, which is a conventional problem; 2) $M > N$, i.e., different subchannels and users. They are investigated in this letter as the channel resource tends to be limited, which means the number of subchannels could be smaller than the number of users.} The details of each case are as follows.
\subsubsection{$N = M$} \textcolor{black}{We consider three situations for this case, $M = N = 2, M = N = 4$ and $ M = N = 8.$}

When $N = M = 2$, there are two users and two subchannels in the optimization problem.  Denote the index of an arbitrary sample by $m$, each input is a  matrix with the size of $2 \times 2$: $R^m = \{r_{ij}^m\}, i,j \in {1,2}$ and the target is a  vector  of the size $1 \times 2: \bar{X}^m = \{x_{1}^m, x_{2}^m\}$. The targets have two kinds of values: $\{1,2\}$ and $\{2,1\}$.
Three FNNs are employed as base models, the top model is also a FNN fed by the stacked outputs of the base learners.

When $N = M = 4$, for each sample $(R^m, \bar{X}^m)$, $R^m$ and $\bar{X}^m$ respectively have the size of $4 \times 4$ and $1 \times 4$.
Three base models are composed of two CNNs and a GRU. The top model is a CNN which is fed by the input with the size $12\times1$ obtained by stacking the outputs of three base models.

When $N = M = 8$, the samples are $(R^m, \bar{X}^m)$ where $R^m$ and $\bar{X}^m$ respectively have the size of $8 \times 8$ and $1 \times 8$.
After comparing the performance of CNNs, GRUs, FNNs and random forest, we chose CNNs as the base and top learners.
The top model is fed by the stacked outputs of three base models, hence the input size of the top model is $1\times24.$

\subsubsection{$M >  N$} \textcolor{black}{We also consider three situations for this case: $M = 4, N = 2$; $M = 8, N = 4$ and $M = 10, N = 5$. With the same procedures, the base models include CNNs, NNs, random forest, and the top model (CNN or NN) is used for models' stacking. }
\vspace{-1em}
\subsection{Permutation}\vspace{-.3em}
\textcolor{black}{After regression, machine learning models usually output decimal values, however, integer outputs are needed as we need to know the exact indices of users. Thus, permutation (or sorting) approach is employed in the end to process the outputs of machine learning models so as to obtain outputs in integer form. More specifically, assume that the target of one sample is $\{3,1,4,2\}$ and the real output is $\{2.15, 0.76, 4.43, 1.55\}$, the real output should be change to integers. The intuitive operations here might be floor,  ceil and round. However, these operations are not appropriate in implementation, as rounding provides the result $\{2, 1, 4, 2\}$, floor renders $\{2, 0, 4, 1\}$ and ceiling leads to $\{3, 1, 5, 2\}$, all of which are unsatisfied outcomes. Permutation sorts the four values based on their relative size and obtains $\{3, 1, 4, 2\}$ as the final prediction result, which is more reasonable and capable to provide better performance in practical applications.}
\vspace{-1em}
\subsection{Accuracy Calculation}\vspace{-0.3em}
In order to provide comprehensive observations of models' capacity and predictive results, 
the proportion of correctly predicted users is defined as the prediction accuracy. More specifically, {assuming} that the number of samples is {$K$}, the accuracy is formulated as follows.
\begin{equation*}\vspace{-.5em}
\text{Accuracy} = \frac{\text{The number of correctly predicted}\ \bar{x}_j^m}{KN}.
\end{equation*}
\begin{table}
	\textcolor{black}{
		\footnotesize
		\renewcommand \arraystretch{1.5} 
		\renewcommand\tabcolsep{2pt}	
		\caption{{Test performance of the cases $N = M = 2, 4, 8$.} }
		\vspace {-0.2cm}
		\begin{center}
			\begin{threeparttable}\scriptsize
				\begin{tabular}{c|c|c|c|c|c}\hline
					{Cases}  & Base model 1 & Base model 2  & Base model 3 & Top Model & Training time \\ \hline
					N = 2 & 99.97\% & 99.91\% & 99.95\% & 99.99\%  & 24.8s \\	\hline
					N = 4 & 97.90\% & 98.41\% & 97.90\% & 98.95\%  & 604.5s\\ \hline
					N = 8 & 89.83\% & 89.83\% & 89.83\% & 90.67\%  & 1218.6s\\ \hline
				\end{tabular}
			\end{threeparttable}
		\end{center}
		\label{tabel:overall1}
		\vspace {-1.5em}
	}
\end{table}
\begin{table}
	{
		\footnotesize
		\renewcommand \arraystretch{1.5} 
		\renewcommand\tabcolsep{0.1pt}	
		\caption{{{Test performance of the cases which have different subchannels and users: $(\text{$M, N$})\!\! = (4, 2), (8, 4), (10, 5)$.} }}
		\begin{center}
			\begin{threeparttable}\scriptsize
				\begin{tabular}{c|c|c|c|c|c}\hline
					\centering{Cases}  & \centering{Base model 1} & \centering{Base model 2}  & \centering{Base model 3} & \centering{Top Model} & Training\ \  time \\ \hline
					$M = 4, N = 2$ & \ 99.51\% & \ 99.26\% &\ 95.20\% &\ 99.73\% &\ 87.47 s\\ 	\hline
					$M = 8, N = 4$ &\ 94.59\% &\ 93.09\% &\ 93.40\% &\ 96.13\% & \ 595.06 s\\ \hline
					$M = 10, N = 5$ &\ 86.03\% &\ 85.74\% &\ 85.09\% &\ 86.69\% &\ 867.85 s\\ \hline
				\end{tabular}
			\end{threeparttable}
		\end{center}
		\label{tabel:overall2}
		\vspace {-2em}
	}
\end{table}
\begin{table}
	\vspace {-2em}
{
	\footnotesize
	\renewcommand \arraystretch{1.5} 
	\renewcommand\tabcolsep{2pt}	
	\caption{{Time consumed (sec) {of 10000 samples} using Algorithm 1 and machine learning methods for 6 cases discussed in this letter. }}
	\begin{center}
		\begin{threeparttable}\scriptsize
			\begin{tabular}{p{0.21\columnwidth}|p{0.1\columnwidth}|p{0.1\columnwidth}|p{0.1\columnwidth}|p{0.1\columnwidth}|p{0.1\columnwidth}|p{0.12\columnwidth}}\hline
				\diagbox{Methods}{Cases} & $M = 2$,\ $N = 2$ & $M = 4$,\ $N = 4$  & $M = 8$,\ $N = 8$ & $M = 4$,\ $N = 2$ & $M = 8$,\ $N = 4$ & $M = 10$,\ $N = 5$\\ \hline
				Algorithm 1 &\ 41.85s &\  48.28s &\ 48.64s &\ 50.12s  &\ 52.66s &\ 56.59s \\	\hline
				ML methods &\ 0.04s &\ 0.93s &\ 2.85s &\ 0.09s &\ 1.26s &\ 2.69s\\
				\hline
			\end{tabular}
		\end{threeparttable}
	\end{center}
	\label{tabel:time}}
	\vspace {-1cm}
\end{table}

%
\section{Test Results}
{The test results are shown in {Tables} \ref{tabel:overall1} and \ref{tabel:overall2}}. From {these tables}, it can be seen that although the prediction accuracy decreases as $N$ increases, the prediction results are satisfactory and the general relationship between the optimization matrix $R$ and the final assignments $X$ can be learned by {the integration of} base and stacking models. When $N =M = 2; N = M = 4; M = 4, N = 2; M = 8, N = 4$, the prediction accuracy is more than 95\% which is better than the {published} results in {literature \cite{lee2018deep}}. {The test accuracy of the cases $N = 4$ and $N = 8$ in literature \cite{lee2018deep}  is respectively 92.76\% and 77.86\% (in both cases the data generation and the training of the deep learning models are done offline in \cite{lee2018deep}), which is lower than the performance of the proposed method about 6\% and 12\%.}

To compare the time efficiency of Algorithms 1 and machine learning methods, we tested the computation time of 10000 samples
{using the same computer (i7-7700 CPU, Nvidia Geforce GTX 1050)}.
The results are shown in Table \ref{tabel:time}, {from which} we can see that the transformation to regression problem which is solved by machine learning techniques largely improves the time efficiency of computation and provides a new perspective of optimization problem solving.

\vspace{-.5em}
\section{Conclusion}\vspace{-.2em}
This letter employs convex optimization based algorithm and machine learning based methods to solve the channel assignment problem. The optimization problem is converted to regression problems for different cases and solved by machine learning methods: CNNs, FNNs, GRUs, random forest and stacking approach. The results demonstrate that machine learning based methods have superiority in improving time efficiency without compromising much prediction accuracy. The future studies may investigate more comprehensive machine learning techniques to achieve higher performance in more complicated scenarios.

\vspace{-.5em}
\bibliographystyle{IEEEtran}
\bibliography{IEEEabrv,MMM}

\begin{thebibliography}{10}
\providecommand{\url}[1]{#1}
\csname url@samestyle\endcsname
\providecommand{\newblock}{\relax}
\providecommand{\bibinfo}[2]{#2}
\providecommand{\BIBentrySTDinterwordspacing}{\spaceskip=0pt\relax}
\providecommand{\BIBentryALTinterwordstretchfactor}{4}
\providecommand{\BIBentryALTinterwordspacing}{\spaceskip=\fontdimen2\font plus
\BIBentryALTinterwordstretchfactor\fontdimen3\font minus
  \fontdimen4\font\relax}
\providecommand{\BIBforeignlanguage}[2]{{%
\expandafter\ifx\csname l@#1\endcsname\relax
\typeout{** WARNING: IEEEtran.bst: No hyphenation pattern has been}%
\typeout{** loaded for the language `#1'. Using the pattern for}%
\typeout{** the default language instead.}%
\else
\language=\csname l@#1\endcsname
\fi
#2}}
\providecommand{\BIBdecl}{\relax}
\BIBdecl

\bibitem{di2016sub}
B.~Di, L.~Song, and Y.~Li, ``Sub-channel assignment, power allocation, and user
  scheduling for non-orthogonal multiple access networks,'' \emph{IEEE Trans.
  Wireless Commun.}, vol.~15, no.~11, pp. 7686--7698, Sep. 2016.

\bibitem{8755300}
M.~{Chen}, U.~{Challita}, W.~{Saad}, C.~{Yin}, and M.~{Debbah}, ``Artificial
  neural networks-based machine learning for wireless networks: {A} tutorial,''
  \emph{IEEE Commun. Surveys Tut.}, vol.~21, no.~4, pp. 3039--3071,
  Fourthquarter 2019.

\bibitem{8752012}
P.~{Dong}, H.~{Zhang}, G.~Y. {Li}, I.~{Gaspar}, and N.~{Naderializadeh}, ``Deep
  {CNN}-based channel estimation for mmwave massive {MIMO} systems,''
  \emph{IEEE J. Sel. Topics Signal Process.}, vol.~13, no.~5, pp. 989--1000,
  Sep. 2019.

\bibitem{chen2019joint}
M.~Chen, Z.~Yang, W.~Saad, C.~Yin, H.~V. Poor, and S.~Cui, ``A joint learning
  and communications framework for federated learning over wireless networks,''
  \emph{arXiv preprint arXiv:1909.07972}, 2019.

\bibitem{yang2019energy}
Z.~Yang, M.~Chen, W.~Saad, C.~S. Hong, and M.~Shikh-Bahaei, ``Energy efficient
  federated learning over wireless communication networks,'' \emph{arXiv
  preprint arXiv:1911.02417}, 2019.

\bibitem{wang2019deep}
Y.~Wang, M.~Chen, Z.~Yang, T.~Luo, and W.~Saad, ``Deep learning for optimal
  deployment of uavs with visible light communications,'' \emph{arXiv preprint
  arXiv:1912.00752}, 2019.

\bibitem{shi2018adaptive}
Q.~Shi, H.-K. Lam, B.~Xiao, and S.-H. Tsai, ``Adaptive pid controller based
  onq-learning algorithm,'' \emph{CAAI Transactions on Intelligence
  Technology}, vol.~3, no.~4, pp. 235--244, 2018.

\bibitem{zappone2018user}
A.~Zappone, L.~Sanguinetti, and M.~Debbah, ``User association and load
  balancing for massive {MIMO} through deep learning,'' in \emph{Proc. IEEE
  Asilomar Conference on Signals, Systems, and Computers}, Pacific Grove, CA,
  USA, Oct. 2018, pp. 1262--1266.

\bibitem{wang2019caching}
Y.~Wang and V.~Friderikos, ``Caching as an image characterization problem using
  deep convolutional neural networks,'' \emph{arXiv preprint arXiv:1907.07263},
  2019.

\bibitem{lee2018deep}
M.~Lee, Y.~Xiong, G.~Yu, and G.~Y. Li, ``Deep neural networks for linear sum
  assignment problems,'' \emph{IEEE Wireless Commun. Lett.}, vol.~7, no.~6, pp.
  962--965, June 2018.

\bibitem{lee2019graph}
M.~Lee, G.~Yu, and G.~Y. Li, ``Graph embedding based wireless link scheduling
  with few training samples,'' \emph{arXiv preprint arXiv:1906.02871}, 2019.

\bibitem{lee2019learning}
------, ``Learning to branch: {A}ccelerating resource allocation in wireless
  networks,'' \emph{arXiv preprint arXiv:1903.01819}, 2019.

\bibitem{7390209}
N.~{Zhang}, J.~{Wang}, G.~{Kang}, and Y.~{Liu}, ``Uplink nonorthogonal multiple
  access in {5G} systems,'' \emph{IEEE Commun. Lett.}, vol.~20, no.~3, pp.
  458--461, Mar. 2016.

\bibitem{8125101}
Z.~{Yang}, W.~{Xu}, Y.~{Pan}, C.~{Pan}, and M.~{Chen}, ``Energy efficient
  resource allocation in machine-to-machine communications with multiple access
  and energy harvesting for {IoT},'' \emph{IEEE Internet Things J.}, vol.~5,
  no.~1, pp. 229--245, Feb. 2018.

\bibitem{boyd2004convex}
S.~Boyd and L.~Vandenberghe, \emph{Convex {O}ptimization}.\hskip 1em plus 0.5em
  minus 0.4em\relax Cambridge University Press, 2004.

\bibitem{bertsekas2009convex}
D.~P. Bertsekas, \emph{{Convex Optimization Theory}}.\hskip 1em plus 0.5em
  minus 0.4em\relax Athena Scientific Belmont, 2009.

\bibitem{langkvist2014review}
M.~L{\"a}ngkvist, L.~Karlsson, and A.~Loutfi, ``A review of unsupervised
  feature learning and deep learning for time-series modeling,'' \emph{Pattern
  Recognition Letters}, vol.~42, pp. 11--24, June 2014.

\bibitem{chung2014empirical}
J.~Chung, C.~Gulcehre, K.~Cho, and Y.~Bengio, ``Empirical evaluation of gated
  recurrent neural networks on sequence modeling,'' \emph{arXiv preprint
  arXiv:1412.3555}, 2014.

\bibitem{1204012}
W.~{Su}, Z.~{Safar}, and K.~J.~R. {Liu}, ``Space-time signal design for
  time-correlated rayleigh fading channels,'' in \emph{Proc. IEEE Int. Conf.
  Commun.}, vol.~5, Anchorage, AK, USA, May 2003, pp. 3175--3179 vol.5.

\bibitem{wolpert1992stacked}
D.~H. Wolpert, ``Stacked generalization,'' \emph{Neural networks}, vol.~5,
  no.~2, pp. 241--259, 1992.

\bibitem{keskar2017improving}
N.~S. Keskar and R.~Socher, ``Improving generalization performance by switching
  from adam to sgd,'' \emph{arXiv preprint arXiv:1712.07628}, 2017.

\end{thebibliography}

\end{document}